\documentclass[letter]{aa} % for the letters 

\usepackage{graphicx}
\usepackage{txfonts}
\usepackage[]{natbib}
\begin{document} 
   \title{Fundamental Vibrational Transitions of HCl Detected
     in CRL~2136 
\fnmsep\thanks{Based on data collected in CRIRES open-time
  program [289.C-5042] at the VLT on Cerro Paranal (Chile),
  which is operated by the European Southern Observatory (ESO).}
\fnmsep\thanks{ Based on data collected in IRCS program
  [S12A-075] at Subaru Telescope on Mauna Kea, Hawaii, operated
  by the National Astronomical Observatory of Japan (NAOJ).}  }

\titlerunning{Fundamental Vibrational Transitions of HCl}
\authorrunning{Goto et al.}

   \author{Miwa Goto\inst{1}, 
           T. Usuda\inst{2}, 
           T. R. Geballe\inst{3},
           K. M. Menten\inst{4},
           N. Indriolo\inst{5},
           D. A. Neufeld\inst{5}}

\institute{Universit\"ats-Sternwarte M\"unchen,
    Ludwig-Maximilians-Universit\"at, Scheinerstr.~1, 81679
     M\"unchen, Germany \email{mgoto@usml.lmu.de}
\and
    Subaru Telescope, 650 North A`ohoku Place, Hilo, HI 96720, USA
\and 
    Gemini Observatory, 670 North A`ohoku Place, Hilo, HI 96720, USA
\and
    Max-Planck-Institut f\"ur Radioastronomie, Auf dem H\"ugel 69,
    53121 Bonn, Germany
\and
    Department of Physics and Astronomy, Johns Hopkins
    University, 3400 N. Charles St., Baltimore, MD 21218, USA }

   \date{} \abstract{}{We would like to understand the chemistry
     of dense clouds and their hot cores more quantitatively by
     obtaining more complete knowledge of the chemical species
     present in them.}  {We have obtained high-resolution
     infrared absorption spectroscopy at 3--4~$\mu$m toward the
     bright infrared source CRL~2136.}  {The fundamental
     vibration-rotation band of HCl has been detected within a
     dense cloud for the first time.  The HCl is probably
     located in the warm compact circumstellar envelope or disk
     of CRL~2136. The fractional abundance of HCl is
     (4.9--8.7)$\times$10$^{-8}$, indicating that approximately
     20\% of the elemental chlorine is in gaseous HCl.  The
     kinetic temperature of the absorbing gas is 250~K, half the
     value determined from infrared spectroscopy of $^{13}$CO
     and H$_2$O. The percentage of chlorine in HCl is
     approximately that expected for gas at this temperature.  The
     reason for the difference in temperatures between the
     various molecular species is unknown.}{}

% http://www.aanda.org/index2.php?option=com_content&task=view&id=170&Itemid=256

%             --- techniques: spectroscopic }

   \maketitle
%-------------------------------------------------------------
\section{Introduction}

Hydrogen chloride (HCl) was first detected in the interstellar
medium by \cite{Blake:1985p43439} in the dense molecular cloud
OMC-1 via its $J=$1-0 rotational emission line. Subsequent
observations in OMC-1 \citep{Schilke:1995p44071}, in Sgr~B2
\citep{Zmuidzinas:1995p43425,Lis:2010p43458,Menten:2011p43254},
in Orion~A and Mon~R2 molecular clouds \citep{Salez:1996p44048},
and in other star forming regions \citep{Peng:2010p44047}
established that HCl is a common constituent of dense clouds, with
typical column densities of (1--20)$\times 10^{13}$~cm$^{-2}$
and fractional abundances with respect to hydrogen of
(1--30)$\times 10^{-10}$. In these clouds chlorine in
HCl comprises less than one percent of elemental chlorine, with
the rest presumably frozen on dust grains
\citep{Schilke:1995p44071,Peng:2010p44047}.

%---
Lines of the fundamental vibration-rotation band of HCl, located
in the 3.2--3.8~$\mu$m region, were first detected in sunspot
spectra by \citet{Hall:1972p44156}, and in the photosphere of an
S-type star by \citet{Ridgway:1984p44159}. Here we report the
first detection of fundamental band lines of HCl in the
interstellar medium, within the dark cloud containing CRL~2136
most probably in the circumstellar disk or envelope of that
object.  CRL~2136 is a massive ($\sim10~M_\odot$), luminous
($\sim 5\times 10^5 L_\odot$) young star in formation at a
distance of 2~kpc \citep{Kastner:1992p31320}. It is bright at
thermal infrared wavelengths (e.g., 3.0~mag in the WISE $W1$
band, 3.4~$\mu$m), and serves as an archetypal line of sight for
the study of gas in dense clouds via absorption
spectroscopy. One of the first detections of interstellar
H$_3^+$ \citep{Geballe:1996p8673} was made toward it near
3.7~$\mu$m, as well the first infrared detection of HF near
2.5~$\mu$m \citep{Indriolo:2013p42312}.  High S/N spectroscopy
of CRL~2136 at other infrared wavelengths has the potential of
yielding detections of additional species.

%-------------------------------------------------------------
\section{Observations}
% CRIRES

The observations reported here were carried out at two
telescopes. On 12 September 2012 spectra of CRL~2136 were
obtained at the VLT on Paranal using the CRIRES spectrograph
\citep{Kaeufl:2004p40429} for DDT program 289.C-5042, whose aim
was to search for the $R$(3,3)$^l$ absorption line of H$_3^+$
(which was not detected). A 0\farcs2 wide slit  (of length
40\arcsec) was
used to achieve a spectral resolution of $R$=100,000. The
adaptive optics system MACAO \citep{Bonnet:2004p44078} was used
to increase the slit throughput by using a nearby visible star
as a wavefront reference. The wavelength regions observed were
centered at  3.576\,$\mu$m, 3.658\,$\mu$m, and
  3.668\,$\mu$m, each meant to cover key H$_3^+$ lines:
$R$(3,3)$^l$, $R$(2,2)$^l$ and the doublet
$R$(1,0)/$R$(1,1)$^u$, respectively. The slit was oriented at
position angle 45\degr~ to avoid the diffuse nebulosity near
CRL~2136.  The telescope was nodded along the slit by
10\arcsec~every two exposures in order to sample the sky
emission spectrum in the same rows as CRL~2136 was observed and
subtract it from the object spectrum.  No spatially extended
  line-emission greater than 1\% of the continuum level
  was detected at the wavelengths of HCl lines in two-dimensional
  spectral images before the sky subtraction, and thus the 
observed absorption lines could only be slightly affected 
by extended line emission. The integration time
was 6 min for each grating setting. A spectroscopic standard
star HR~6879 \citep[B9.5\,III; $L=$1.707~mag;][]
{Bouchet:1991p9616} was observed to allow accurate removal of
telluric absorption lines. Other standard calibration data,
including spectroscopic flat field and dark current images, were
obtained at the end of the night.

% IRCS

Spectra of CRL~2136 in the 3--4~$\mu$m region were also obtained
at the Subaru Telescope on Mauna Kea on 15 June 2012 with the
cross-dispersing spectrograph IRCS in the open-time program
S12A-075. A 0\farcs14$\times$6\farcs7 slit was used to record
the spectra at $R$=20,000. The slit was oriented at position
angle 0\degr. Two grating settings were used to cover selected
wavelength intervals from 3.25~$\mu$m to 4.13~$\mu$m.  The
integration times for the two grating settings were 6 min and 12
min. The telescope was nodded by 3\farcs4 along the slit to
sample the sky emission. A spectroscopic standard star HR~6556
(A5\,III; $L=$1.61~mag\footnote{From UKIRT $L^\prime$ $M^\prime$
  standards.
  http://www.jach.hawaii.edu/UKIRT/astronomy/calib/phot\_cal/lm\_stds.html})
was observed before the observation of CRL~2136.  Other
calibration data were obtained on the morning after the
observation. Another observation was performed with IRCS on UT
12 May 2013 in the open-time program S13A-077, using three
grating settings which provided complete coverage from
3.20~$\mu$m to 4.15~$\mu$m continuously. In this case the slit
was oriented at position angle 45\degr~and the integration times
were 6 to 24 min. The spectroscopic standard star HR~7001 (Vega;
A0\,V; $L=$0.00~mag\footnote{From UKIRT $L^\prime$ $M^\prime$
  standards.
  http://www.jach.hawaii.edu/UKIRT/astronomy/calib/phot\_cal/lm\_stds.html}
was observed before the science observation for the purpose of
telluric line removal.

The CRIRES data were reduced using the CRIRES pipeline
recipe\footnote{http://www.eso.org/sci/software/pipelines/crire/crire-pipe-recipes.html}
ver. 2.3.1 on
Gasgano\footnote{http://www.eso.org/sci/software/gasgano/} and
EsoRex\footnote{http://www.eso.org/sci/software/cpl/esorex.html}
platforms provided by the European Southern Observatory
(ESO). The pipeline reduction included correction of the
detector linearity, subtraction of the sky emission,
normalization of the detector pixel responses, and extraction of
one-dimensional spectra. The spectra obtained at the two
locations of CRL~2136 on the slit were extracted separately to
avoid degrading the spectral resolution of the combined spectrum
due to the distortion of the slit image and the different
degrees of defocusing at the different locations on the slit.
Wavelength calibration was achieved by matching the observed
telluric absorption lines to the model atmospheric transmission
spectrum calculated by LBLRTM \citep{Clough:2005p38894}. In
terms of velocity the uncertainty in the calibration is
typically less than 1~km\,s$^{-1}$. The telluric absorption
lines were removed by dividing the spectra of CRL~2136 by those
of the standard stars. Slight mismatches of the wavelength,
spectral resolution, and optical depth of the absorption lines
between the object and the standard star spectra were manually
corrected to the extent possible. The spectra extracted at the
two locations on the slit were added together after the
wavelength calibration and the removal of the telluric lines
were performed individually. The calibrated wavelengths were
converted to the velocities with respect to the local standard
of rest. The data obtained by IRCS were reduced in a similar
manner except that the spectral extraction was done with the
IRAF aperture extraction package\footnote{IRAF is distributed by
  the National Optical Astronomy Observatories, which are
  operated by the Association of Universities for Research in
  Astronomy, Inc., under cooperative agreement with the National
  Science Foundation.}. Wavelength calibration was performed in
the same way as for the CRIRES data; the uncertainty corresponds
to a few~km\,s$^{-1}$. The IRCS spectra from 2012 and 2013 were
coadded to increase the signal-to-noise ratio in those instances
where the same lines were observed.

%-------------------------------------------------------------
%                 A figure as large as the width of the column
   \begin{figure}
   \centering
   \includegraphics[height=\hsize, angle=-90]{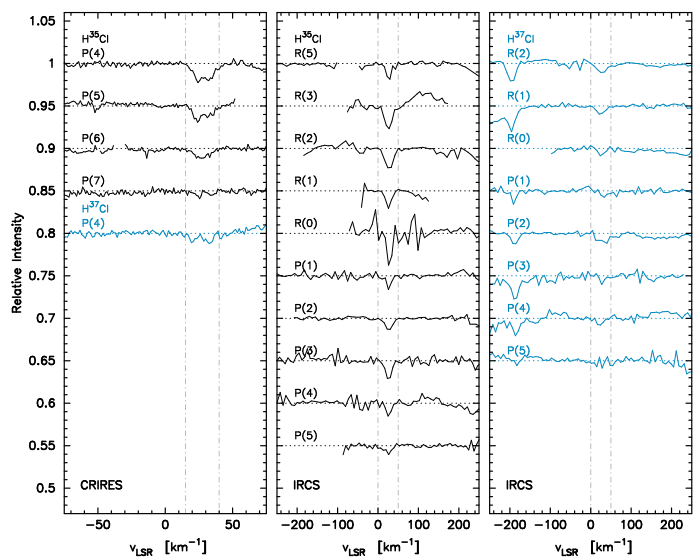}
   \caption{HCl fundamental vibrational transitions near
     3.5~$\mu$m. Left: spectra obtained by CRIRES at the
     VLT. Middle and right: spectra obtained by IRCS at the
     Subaru Telescope. The intervals used to calculate the
     equivalent widths are marked by dot-dashed lines.  A bump
     on the long wavelength shoulder of H$^{35}$Cl $R$(3) IRCS
     spectrum is a residual of a strong and improperly cancelled
     telluric methane absorption line. \label{f1}}
   \end{figure}

%-------------------------------------------------------------
   \begin{figure}
   \centering

   \includegraphics[height=\hsize,angle=-90]{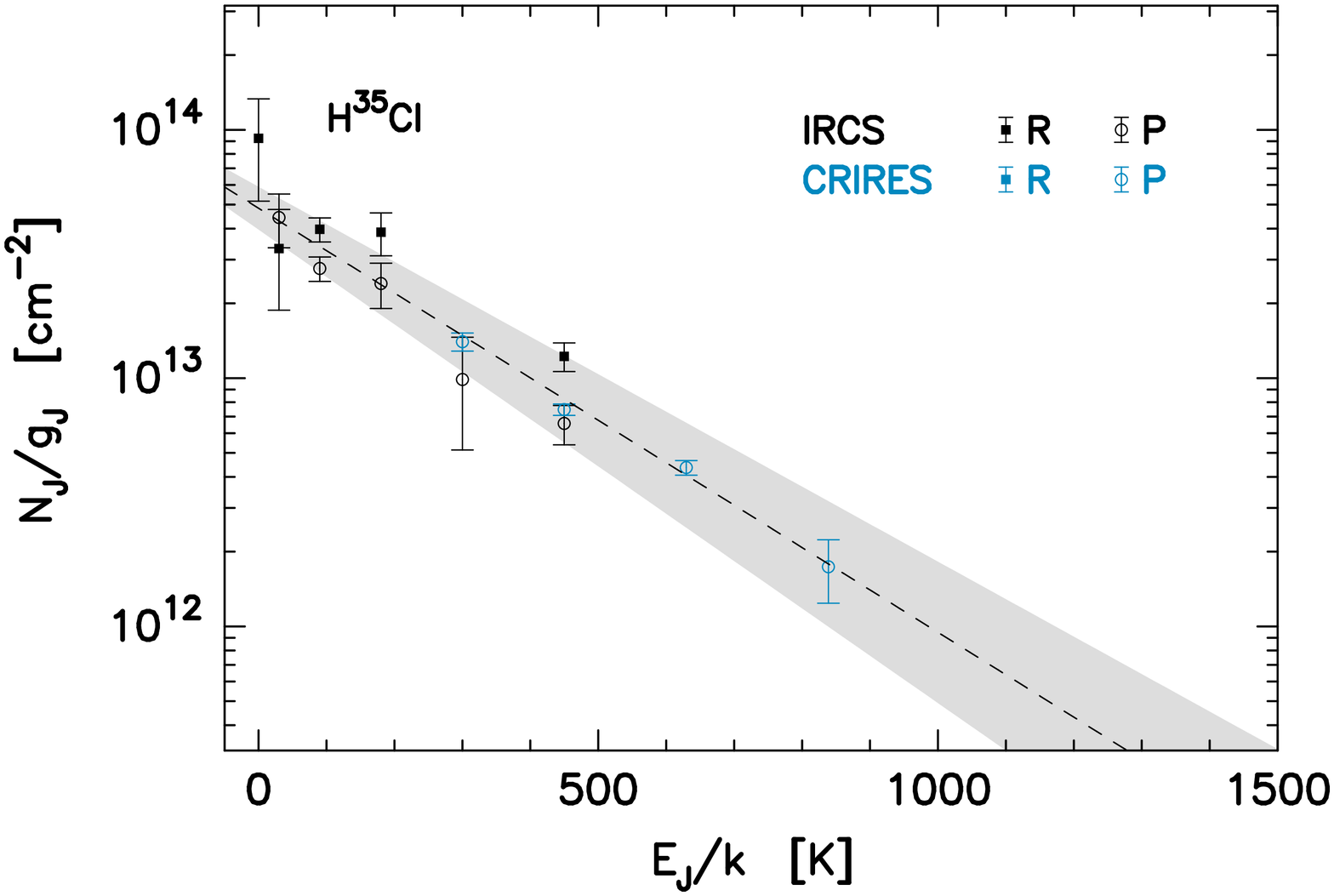}
   \includegraphics[height=\hsize,angle=-90]{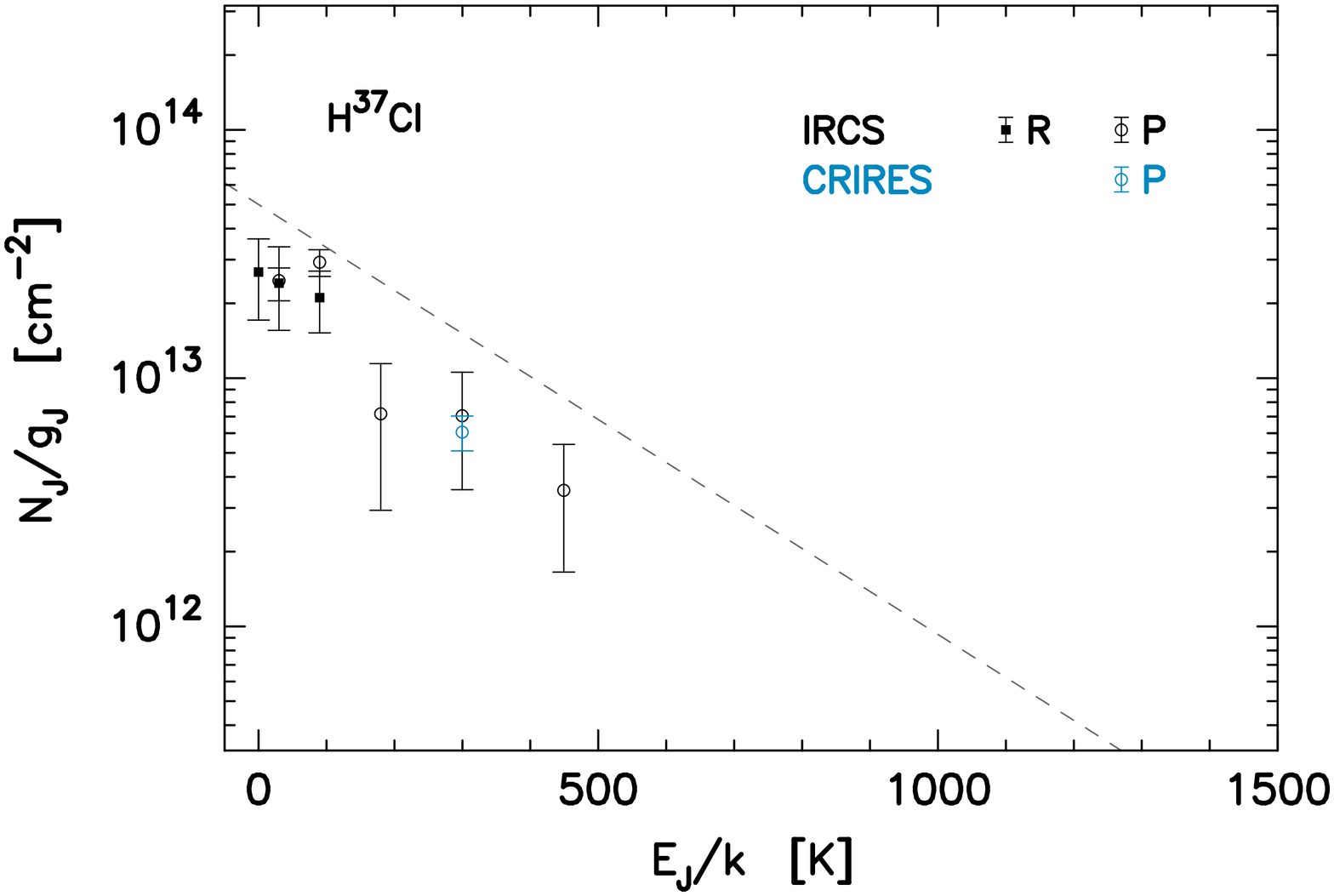}
      \caption{Rotational population diagrams of H$^{35}$Cl and
        H$^{37}$Cl. $R$-branch lines (filled squares) and $P$
        branch lines (open circles) are shown separately. The
        level column densities of H$^{35}$Cl were fit by a
        dashed line; the same fit is shown for H$^{37}$Cl. 
          The shaded area depicts the range of LTE fitting with
          the excitation temperatures and the total column
          densities deviating by up to $\pm$3~$\sigma$ from the best-fit
          values. \label{f2}}
   \end{figure}
%
%-------------------------------------------------------------
% \clearpage
%============
% table 1
%============
\begin{table*}
\begin{center}

\caption{HCl lines detected.\label{t1}}
\begin{tabular}{l c c rr rr cr cr}
\hline \hline
%---
& &                       &          &                  &       &                & \multicolumn{2}{c}{CRIRES}                        & \multicolumn{2}{c}{IRCS}\\
& & $\lambda_{JJ^\prime}$ &  \multicolumn{1}{c}{$E_J/k$} & $A_{J^\prime J}$ & $g_J$ & $g_{J^\prime}$ & $W_\lambda(JJ^\prime)$ & \multicolumn{1}{c}{$N_J$}                  & $ W_\lambda(JJ^\prime)$ & \multicolumn{1}{c}{$N_J$}  \\
& &       [$\mu$m]        & \multicolumn{1}{c}{[K]}    &    [s$^{-1}$]    &       &                & [10$^{-6}$\,$\mu$m]      & [10$^{14}$\,cm$^{-2}$] &   [10$^{-6}$\,$\mu$m]     & [10$^{14}$\,cm$^{-2}$] \\
\hline
H$^{35}$Cl  & $R$(5) & 3.3355 &   449.95 & 17.72 &     88 &    104 &               &               & 3.7 $\pm$ 0.5 & 10.8 $\pm$ 1.4 \\
            & $R$(3) & 3.3746 &   180.14 & 17.39 &     56 &     72 &               &               & 6.9 $\pm$ 1.6 & 17.9 $\pm$ 4.2 \\
            & $R$(2) & 3.3957 &    90.10 & 16.90 &     40 &     56 &               &               & 6.6 $\pm$ 0.7 & 15.9 $\pm$ 1.8 \\
            & $R$(1) & 3.4178 &    30.04 & 15.89 &     24 &     40 &               &               & 3.8 $\pm$ 2.6 & 8.0 $\pm$ 5.4 \\
            & $R$(0) & 3.4409 &     0.00 & 13.32 &      8 &     24 &               &               & 5.5 $\pm$ 2.4 & 7.4 $\pm$ 3.3 \\
            & $P$(1) & 3.4903 &    30.04 & 40.33 &     24 &      8 &               &               & 2.8 $\pm$ 0.7 & 10.6 $\pm$ 2.5 \\
            & $P$(2) & 3.5166 &    90.10 & 26.97 &     40 &     24 &               &               & 3.6 $\pm$ 0.4 & 11.1 $\pm$ 1.1 \\
            & $P$(3) & 3.5441 &   180.14 & 24.33 &     56 &     40 &               &               & 4.9 $\pm$ 1.0 & 13.5 $\pm$ 2.8 \\
            & $P$(4) & 3.5728 &   300.12 & 23.20 &     72 &     56 & 3.9 $\pm$ 0.3 & 10.1 $\pm$ 0.8 & 2.8 $\pm$ 1.3 & 7.1 $\pm$ 3.4 \\
            & $P$(5) & 3.6026 &   449.95 & 22.57 &     88 &     72 & 2.7 $\pm$ 0.1 & 6.6 $\pm$ 0.3 & 2.4 $\pm$ 0.4 & 5.8 $\pm$ 1.0 \\
            & $P$(6) & 3.6337 &   629.54 & 22.14 &    104 &     88 & 2.0 $\pm$ 0.1 & 4.5 $\pm$ 0.3 &               &               \\
            & $P$(7) & 3.6660 &   838.80 & 21.83 &    120 &    104 & 0.9 $\pm$ 0.3 & 2.1 $\pm$ 0.6 &               &               \\
\hline
H$^{37}$Cl  & $R$(2) & 3.3982 &    89.96 & 16.85 &     40 &     56 &               &               & 3.5 $\pm$ 1.0 & 8.4 $\pm$ 2.4 \\
            & $R$(1) & 3.4203 &    29.99 & 15.84 &     24 &     40 &               &               & 2.8 $\pm$ 0.4 & 5.8 $\pm$ 0.9 \\
            & $R$(0) & 3.4434 &     0.00 & 13.28 &      8 &     24 &               &               & 1.6 $\pm$ 0.6 & 2.1 $\pm$ 0.8 \\
            & $P$(1) & 3.4928 &    29.99 & 40.21 &     24 &      8 &               &               & 1.6 $\pm$ 0.6 & 5.9 $\pm$ 2.2 \\
            & $P$(2) & 3.5192 &    89.96 & 26.89 &     40 &     24 &               &               & 3.8 $\pm$ 0.5 & 11.7 $\pm$ 1.4 \\
            & $P$(3) & 3.5467 &   179.87 & 24.26 &     56 &     40 &               &               & 1.5 $\pm$ 0.9 & 4.0 $\pm$ 2.4 \\
            & $P$(4) & 3.5753 &   299.67 & 23.13 &     72 &     56 & 1.7 $\pm$ 0.3 & 4.4 $\pm$ 0.7 & 2.0 $\pm$ 1.0 & 5.1 $\pm$ 2.5 \\
            & $P$(5) & 3.6051 &   449.27 & 22.50 &     88 &     72 &               &               & 1.3 $\pm$ 0.7 & 3.1 $\pm$ 1.7 \\
\hline
\end{tabular}
\tablefoot{The wavelength $\lambda_{JJ^\prime}$ of the
  transitions, the energy of the lower level $E_J/k$, the
  spontaneous emission coefficient $A_{J^\prime J}$, and the
  statistical weights of the lower and upper levels $g_J$ and
  $g_{J^\prime}$, respectively, are taken from the HITRAN
  database \citep{Rothman:2009p44133}.}
\end{center}
\end{table*}
5 \normalsize

% \input{t1.tex}
% \input{t2.tex}
%-------------------------------------------------------------
\section{Results}

Absorption by the v=1-0 $P$(4)--$P$(7) lines of H$^{35}$Cl and the
$P$(4) line of H$^{37}$Cl are clearly detected in the CRIRES
spectra (left panel in Fig.~\ref{f1}) and are centered at
$v_{\rm LSR}$ $\approx +$28~km\,s$^{-1}$. The lines have full
widths at half maximum of $\approx$15~km\,s$^{-1}$. Although
some of the profiles appear to be double-peaked, the S/N is low
and the peaks do not match well in velocity from line to
line. Further investigation will require spectra of higher
accuracy. The $R$(5), $R$(3)--$R$(0), and $P$(1)--$P$(5)
absorption lines of H$^{35}$Cl and the $R$(2)--$R$(0) and
$P$(1)--$P$(5) absorption lines of H$^{37}$Cl were detected in
the IRCS spectra at the same velocity as the CRIRES detections
(middle and right panels in Fig.~\ref{f1}). The lines as viewed
by the IRCS are barely resolved.

The line equivalent widths $W_{\lambda}(JJ^\prime)$ have been
measured by integrating the absorptions over the velocity
intervals $+$15$\rightarrow$$+$40~km\,s$^{-1}$ for the CRIRES
spectra and 0$\rightarrow$$+$50~km\,s$^{-1}$ for the IRCS
spectra. The uncertainties in equivalent widths were
  conservatively estimated from the standard deviations of the
  continuum level in nearby wavelength intervals (where the
  dispersion of continuum values is relatively large) multiplied
  by the wavelength intervals of the integration divided by the
  square root of the number of the data points that sample the
  line absorption profile.

The column densities of the lower levels of the transitions
$N_J$ have been calculated from $W_{\lambda}(JJ^\prime)$ using
\[
\frac{W_{\lambda}(JJ^\prime)}{\lambda} = 
\frac{\lambda^3}{8\,\pi c}\frac{g_{J^\prime}}{g_J} A_{J^\prime J} N_J,
\]
\noindent where $\lambda$ is the wavelength of the transition,
$g_{J^\prime}$ and $g_J$ the statistical weights of the upper
and the lower states, and $A_{J^\prime J}$ the spontaneous
emission coefficient of the transition.  The $A$ coefficients
and statistical weights are taken from the HITRAN database
\citep{Rothman:2009p44133}, retrieved through the web interface
HITRAN on the Web\footnote{http://hitran.iao.ru/}. The results
are summarized in Table~\ref{t1}.

For each HCl isotopomer the level column densities divided by
the statistical weights of the lower levels are plotted in
Fig.~\ref{f2} as a function of lower level energy. For
$N$(H$^{35}$Cl) the column densities obtained from the CRIRES
and the IRCS spectra can be simultaneously fit by a single
straight line without adjustment between the two data sets. The
total column density is given by  $N_0 Q(T)$, where $N_0$
is the ordinate intercept of the line and $Q(T) =
\displaystyle\sum_J g_J \, \exp(-E_J/kT)$ is the partition
function. The latter was calculated up to $J$=26 using the
HITRAN database and the excitation temperature determined from
the slope of the line. The resulting total column density
$N$(H$^{35}$Cl) is  (6.7$\pm$0.4)$\times
10^{15}$~cm$^{-2}$, and the excitation temperature is 
  254$\pm$10~K. The uncertainties include the fitting
uncertainties only. The same treatment cannot be applied to
H$^{37}$Cl because of insufficient sampling of the level column
densities.  However, the isotope ratio [H$^{35}$Cl]/[H$^{37}$Cl]
can be estimated by comparing the level column densities at
$J$=4, as the $P$(4) line was detected for both H$^{35}$Cl and
H$^{37}$Cl by CRIRES.  Assuming identical excitation
temperatures, which is justified by the similar slopes of the
population diagrams, [H$^{35}$Cl]/[H$^{37}$Cl] =
2.3$\pm$0.4. The total column density of HCl (both isotopomers)
is then (9.6 $\pm$ 0.7)$\times 10^{15}$~cm$^{-2}$.

%-------------------------------------------------------------
   \begin{figure}
   \centering

   \includegraphics[width=\hsize,angle=0]{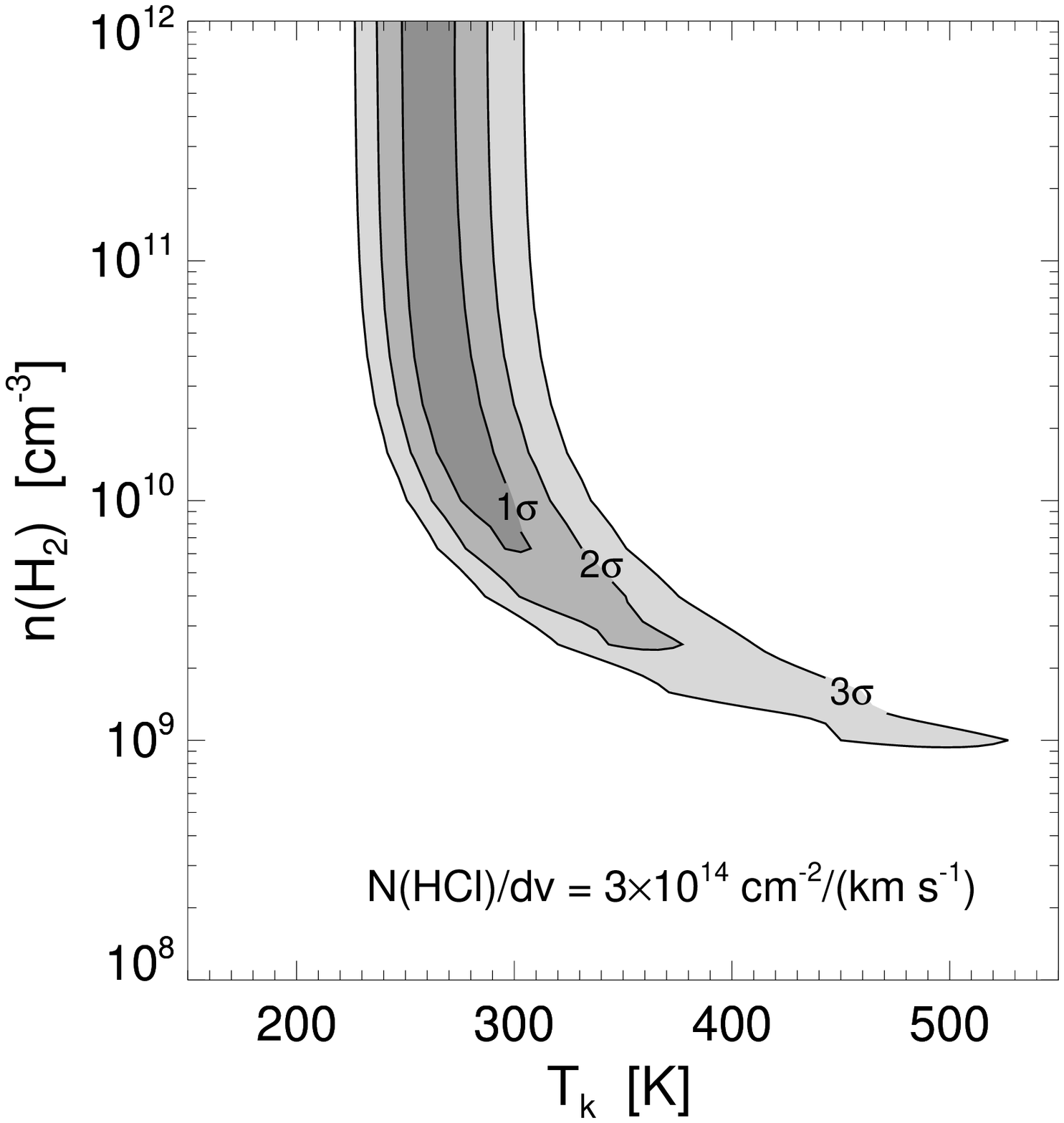}
      \caption{Collisional excitation calculation of HCl in
        CRL~2136. The population diagram of H$^{35}$Cl in
        Fig.~\ref{f2} is fit by the models with varying gas
        temperature and density, and the goodness of the fit is
        presented in terms of the standard deviation. Line
        radiation transfer is included in the model with the
        column density over velocity $3\times
        10^{14}$~cm$^{-2}$/(km\,s$^{-1}$), which is consistent
        with the observed velocity dispersion in the
        infrared.          \label{f3}}
% The continuum radiation pumping is not  included.}
   \end{figure}
%
%-------------------------------------------------------------
\section{Discussion}

Infrared spectroscopy of $^{13}$CO by
\citet{Mitchell:1990p42796} toward CRL~2136 indicates that the
line of sight to the star passes through clouds with two
distinct temperatures, cold gas (17~K) and warm gas (580~K),
with two-thirds of the total column density residing in the warm
gas. The difference between the latter temperature and that
derived for HCl is discussed below. The density of the warm gas
is high, as is shown by $^{13}$CO levels being populated up to
at least $J=$24 (implying $n_{\rm crit} > 5\times
10^7$~cm$^{-2}$). Gas at these temperatures and densities is
likely to be directly involved in the star formation process,
and is probably in a circumstellar envelope or disk. The
intensity-weighted mean of the centers of the HCl lines (as
measured by CRIRES) is $+$27.5$\pm$0.5~km\,s$^{-1}$, which is
identical within the uncertainties to that of the $^{13}$CO
lines \citep[$+$26.5$\pm$2.8~km~s$^{-1}$][]{Mitchell:1990p42796}
and close to that of HF
\citep[$+$23.7--$+$24.8~km~s$^{-1}$][]{Indriolo:2013p42312}. The
above velocities are red-shifted from the systemic velocity of
CRL~2136 derived from the rotational lines of CO
\citep[$+$22~km\,s$^{-1}$;][]{Kastner:1994p31307} and CS
\citep[$+$22.8~km\,s$^{-1}$;][]{vanderTak:2000p31274}, but agree
well with the radial velocity of the water masers
\citep[$+$26.5~km\,s$^{-1}$; ][]{Menten:2004p31266}. The
near-match of the velocities reinforces the above conclusion
that the gas absorption lines are formed in the envelope or disk
of CRL~2136. However, the small but significant positive
velocity shift relative to the pure rotational emission lines
suggest that the infrared-absorbing gas is infalling or
contracting.

The total column density of HCl may be compared with that of
molecular hydrogen, whose v=1-0 $S$(0) line at 2.223~$\mu$m was
detected in absorption by
\citet{Kulesa:2002p42869}. \citet{Indriolo:2013p42312} have
estimated the H$_2$ column density in the warm component,
assuming that [CO]/[H$_2$] is constant along the line of sight
and that the excitation temperature of H$_2$ is same as that
determined by \citet{Mitchell:1990p42796} from the $^{13}$CO
v=1-0 lines. They obtain $N$(H$_2$) = (5.5--9.8)$\times
10^{22}$~cm$^{-2}$, and therefore $N$(HCl)/$N$(H) = 
  (4.9--8.7)$\times 10^{-8}$.  Assuming the solar abundance of
chlorine, $3.2\times 10^{-7}$ \citep{Asplund:2009p44201} in
CRL~2136, approximately 20\% (15--27\%) of the chlorine in
the warm CRL~2136 envelope is in the form of gaseous HCl.

%GOTo-SAN - NEW REFERENCE IN FOLLOWING PARAGRAPH; ADD TO BIBLIOGRAPHY

In diffuse clouds chlorine is singly ionized, as the ionization
potential of the neutral form (13.0~eV) is slightly lower than
that of atomic hydrogen; Cl$^{+}$ can directly react with H$_2$,
leading to additional reactions that produce neutral HCl
\citep[see][for details]{Monje:2013p44198}. In dense clouds,
where virtually no photoionization photons are available, the
formation of HCl is initiated by the proton hop reaction of Cl
with H$_3^+$, and subsequent reaction of the cation with H$_2$,

\begin{eqnarray*}
\mathrm{Cl    + H_3^+ } &\rightarrow& \mathrm{HCl^+   + H_2} \\
\mathrm{HCl^+ + H_2   } &\rightarrow& \mathrm{H_2Cl^+ + H}.
\end{eqnarray*}
\noindent As the ionization fraction is very low, 
dissociative recombination of H$_2$Cl$^+$ on electrons 
is not important. Instead H$_2$Cl$^+$ participates in a second 
proton hop reaction with either CO or H$_2$O,
\begin{eqnarray*}
\mathrm{H_2Cl^+ + CO  } &\rightarrow& \mathrm{HCl     + HCO^+,~~~or} \\
\mathrm{H_2Cl^+ + H_2O} &\rightarrow& \mathrm{HCl     + H_3O^+,}
\end{eqnarray*}
\noindent to form HCl. In CRL~2136 the gas temperature is sufficiently high 
that the only slightly endothermic neutral-neutal reaction ($\Delta E=518$~K)
\[
\mathrm{Cl + H_2 }\rightarrow \mathrm{HCl + H}
\]
\noindent can also contribute. 

Destruction of neutral HCl is expected to be inefficient in
dense clouds.  Photodissociation is slow because of the lack of
UV photons. Cations as reaction partners to promote more
efficient ion-neutral reactions are also in short
supply. Reactions with H$_3^+$ will occur, but will lead to
another cycle of HCl formation. In the circumstellar envelope of
CRL~2136, where the gas is shielded from photodissociating UV
photons, yet the temperature is both high enough for
neutral-neutral reactions to occur and to keep HCl from
accumulating on dust grains, several tens of percent of
elemental chlorine are expected to be present in HCl, with most
of the rest in neutral atomic form \citep[][their
  Fig.~13]{Neufeld:2009p43447}. Our result is consistent with
this.

%-------------------------------------------------------------
%============
% table 2
%============
\begin{table*}
\begin{center}

\caption{Excitation temperature of molecules toward CRL~2136 observed in infrared absorption spectroscopy of vibrational transitions.\label{t2}}
\begin{tabular}{l c r l c l c}
\hline \hline
%---
           & \multicolumn{1}{c}{Transition}& \multicolumn{1}{c}{Wavelength} & \multicolumn{1}{c}{Excitation} & \multicolumn{1}{c}{Telescope/} & \multicolumn{1}{c}{Spectral}& \multicolumn{1}{c}{Reference} \\
           &                     &    & \multicolumn{1}{c}{Temperature [K]} & \multicolumn{1}{c}{Instrument} & \multicolumn{1}{c}{resolution}   &  \\
\hline   
HCl        & $\nu_1$             & 3.5~$\mu$m   & \phantom{0000}254$\pm$10          & VLT/CRIRES   & $R$=100,000  & 1\\
CO$_2$     & $\nu_1$             & 15.0~$\mu$m  & \phantom{0000}300$^{+100}_{-100}$ & ISO/SWS      & $R$=1,500    & 2\\
SO$_2$     & $\nu_3$             & 7.4~$\mu$m   & \phantom{0000}350$^{+100}_{-50}$  & ISO/SWS      & $R$=2,500    & 3\\
H$_2$O     & $\nu_2$             & 6.2~$\mu$m   & \phantom{0000}500$^{+250}_{-150}$ & ISO/SWS      & $R$=1,500    & 4\\
H$_2$O     & $\nu_1,\nu_2,\nu_3$ & 2.5~$\mu$m   & \phantom{0000}506$\pm$25          & VLT/CRIRES   & $R$=100,000  & 5\\
$^{13}$CO  & $\nu_1$             & 4.7~$\mu$m   & \phantom{0000}580$^{+60}_{-50}$   & CFHT/FTS     & $R$=37,000   & 6\\
HCN        & $\nu_2$             & 14.0~$\mu$m  & \phantom{0000}600$^{+75}_{-50}$   & ISO/SWS      & $R$=1,800    & 7\\
C$_2$H$_2$ & $\nu_5$             & 13.7~$\mu$m  & \phantom{0000}800$^{+150}_{-100}$ & ISO/SWS      & $R$=1,800    & 7\\
\hline
\end{tabular}
\tablebib{1~Present
  study. 2~\cite{Boonman:2003p45504}. 3~\cite{Keane:2001p44920}.
  4~\cite{Boonman:2003p44230}. 5~\cite{Indriolo:2013p44653}. 6~\cite{Mitchell:1990p42796}.
  7~\cite{Lahuis:2000p44642}.}
\end{center}
\end{table*}
\normalsize
%-------------------------------------------------------------

A collisional excitation and line radiation transfer model
\citep{Neufeld:2012p41792} has been used to estimate the gas
density and the kinetic temperature (Fig.~\ref{f3}). The
parameters that best reproduce the observed population diagram
are $T_k=$250~K and $n$(H$_2$)$>10^9$~cm$^{-3}$ for a 
  column density over velocity
$N$(HCl)$=3\times10^{14}$~cm$^{-2}$/(km\,s$^{-1}$) that matches
the observed line widths. The accurate straight line fit to the
population diagram (Fig.~\ref{f2}) implies that the HCl is in
LTE. However, as mentioned earlier, this value of the kinetic
temperature is considerably lower than the temperature of
$\sim$500~K determined from absorption spectroscopy of $^{13}$CO
\citep{Mitchell:1990p42796} and of H$_2$O
\citep[][]{Boonman:2003p44230,Indriolo:2013p44653}, whose lines
appear to be formed in gas of similar density to the HCl
lines. The reason for this difference in temperature is
unclear. The high gas density required to populate the observed
rotational levels of HCl clearly excludes the possibility that
most of the observed HCl is in cold gas in the outer part of the
dense cloud \citep[17~K;][]{Mitchell:1990p42796}.

In terms of temperature, HCl is not the only outlier, but
neither is it consistent with all other temperature
determinations. Six other molecules in the line of sight, all
observed using infrared absorption spectroscopy, have excitation
temperatures ranging from 300~K to 800~K, which cannot be
reconciled by the reported uncertainties (Table~\ref{t2}).
Moreover, in this regard CRL~2136 may not be an exception. The
excitation temperatures of different molecules have a
substantial dispersion in the hot cores of six high-mass star
forming regions studied by \cite{Keane:2001p44920} with some
sightlines producing values from 180~K to 800~K (see their
Table~1). A cautionary note for CRL~2136 is that the majority of
the observations in Table~\ref{t2} were made by the Infrared
Space Observatory's moderate resolution SWS spectrometer, for
which vibration-rotation bands are not resolved and therefore
appear broad and smooth.  Excitation temperatures for these data
were estimated by comparing model spectra convolved to the
spectral resolution of SWS ($R$=1500--2500) with the observed
spectrum. For NGC~7538 IRS~1 \citet{Knez:2009p45835} have
pointed out that, by reducing the Doppler broadening parameter
$b$ from 5~km\,s$^{-1}$ of \cite{Lahuis:2000p44642} to
1~km\,s$^{-1}$, which is indicated by their much higher
resolution spectra, the best-fit excitation temperature
decreased from 800~K to 190--230~K for C$_2$H$_2$ and 600~K to
250--450~K for HCN. Furthermore, they point out that bands of
other molecules overlap with those of the dominant species and,
if not identified, also affect temperature determinations for
these species. 

The band structures of HCl in the present study and of H$_2$O by
\citet{Indriolo:2013p44653} are spectrally resolved, and each
observed rotational line is velocity-resolved. Thus, the above
cautionary note probably does not apply to our data. It is
possible that H$_2$O observed at high resolution by
\citet{Indriolo:2013p44653} mainly traces the hottest gas where
the water ice evaporates most rapidly. The excitation
temperature of HCl may be regulated by the endothermic formation
reaction of HCl that sharply kicks in at $T\sim$250~K. However,
until more detailed observations of these and additional
molecular species are available the disagreement in the
excitation temperatures of the various species remains an open
issue.

%-------------------------------------------------------------
\begin{acknowledgements}

  We thank all the staff and crew of the VLT and Subaru
  Telescope for their valuable assistance in obtaining the
  data. We appreciate the hospitality of the Chilean and
  Hawaiian community that made the research presented here
  possible. We appreciate the constructive criticisms of the
  anonymous referee that improved the manuscript. M.G. is
  supported by DFG grant GO 1927/3-1. T.R.G. is supported by the
  Gemini Observatory, which is operated by the Association of
  Universities for Research in Astronomy, Inc., on behalf of the
  international Gemini partnership of Argentina, Australia,
  Brazil, Canada, Chile, and the United States of America.
\end{acknowledgements}

%-------------------------------------------------------------
\bibliographystyle{aa} % style aa.bst
\bibliography{aa} % your references Yourfile.bib
%-------------------------------------------------------------
Note added in proof:

The isotopomer ratio [H$^{35}$Cl]/[H$^{37}$Cl] derived based on
the equivalent widths of $P$(4) lines of CRIRES data might be
affected by a CH doublet at 3.575~$\mu$m that overlaps with
the H$^{37}$Cl $P$(4) line. The combined abundance of HCl may be
affected consequently. The possibility is remote, though, since
CH traces diffuse gas.

%_____________________________________________________________

\end{document}